\documentclass[apjl,a4paper,12pt,useAMS]{emulateapj}
\usepackage{amssymb}
\usepackage{graphicx}

\begin{document}
\title{Revisiting the dispersion measure of fast radio bursts associated with gamma-ray burst afterglows}
\author{Yun-Wei Yu}

\altaffiltext{1}{Institute of Astrophysics, Central China Normal
University, Wuhan 430079, China, {yuyw@mail.ccnu.edu.cn}}

\begin{abstract}
Some fast radio bursts (FRBs) are expected to be associated with the
afterglow emission of gamma-ray bursts (GRBs), while a short-lived,
supermassive neutron star (NS) forms during the GRBs. I investigate
the possible contributions to the dispersion measure (DM) of the
FRBs from the GRB ejecta and the wind blown from the precollapsing
NS. On the one hand, sometimes an internal X-ray plateau afterglow
could be produced by the NS wind, which indicates that a great
number of electron-positron pairs are carried by the wind. If the
pair-generation radius satisfies a somewhat rigorous condition, the
relativistic and dense wind would contribute a high DM to the
associated FRB, which can be comparable to and even exceed the DM
contributed by the intergalactic medium. On the other hand, if the
wind only carries a Goldreich-Julian particle flux, its DM
contribution would become negligible; meanwhile, the internal
plateau afterglow would not appear. Alternatively, the FRB should be
associated with a GRB afterglow produced by the GRB external shock,
i.e., an energy-injection-caused shallow-decay afterglow or a normal
single-power-law afterglow if the impulsive energy release of the
GRB is high enough. In the latter case, the DM contributed by the
high-mass GRB ejecta could be substantially important, in
particular, for an environment of main-sequence stellar wind. In
summary, a careful assessment on the various DM contributors could
be required for the cosmological application of the expected FRB-GRB
association. The future DM measurements of GRB-associated FRBs could
provide a constraint on the physics of NS winds.
\end{abstract}

\section{Introduction\label{sec:intro}}
Fast radio bursts (FRBs) are newly discovered radio transient
sources; they have a typical duration of a few milliseconds and a
flux of a few to a few tens of Jansky at $\sim1$ GHz (Lorimer et al.
2007; Thornton et al. 2013). Due to the low angular resolutions of
the radio surveys for FRBs, no counterpart in other bands has been
reported to be associated with them. In view of their anomalously
high dispersion measures (DMs; $\sim 500-1000\rm ~cm^{-3}pc$)
coupled with their high Galactic latitudes, FRBs are increasingly
suggested to have cosmological distances (Thornton et al. 2013). The
corresponding redshifts are inferred to $z\sim 0.5-1$ by ascribing
the DMs to the host galaxies and the intergalactic medium (IGM;
Thornton et al. 2013). Consequently, the peak radio luminosity is
estimated to be $\sim10^{42-43}~\rm erg ~s^{-1}$ and the total
energy release is $\sim10^{39-40} \rm erg$. Based on such an energy
scale and the millisecond duration, some cosmological FRB models
have been proposed, such as hyperflares of soft gamma-ray repeaters
(Popov \& Postnov 2007), collapses of supra-massive neutron stars
(NSs) to black holes at several thousand to million years old
(Falcke \& Rezzolla 2014), mergers of double NSs (Totani 2013) or
binary white dwarfs (Kashiyama et al. 2013), and synchrotron maser
emission from relativistic, magnetized shocks due to magnetar flares
(Lyubarsky 2014).

It is believed that a supermassive NS could form during some
gamma-ray bursts (GRBs; in particular, the short-duration ones) and
subsequently collapse into a black hole after hundreds to thousands
of seconds from its birth. Therefore, following Falcke \& Rezzolla
(2014), Zhang (2014) proposed a possible connection between a small
fraction\footnote{The event rate of FRBs is considered to be much
higher than the GRB rate (Thornton et al. 2013).} of FRBs and GRBs,
although no such association (even a positional coincidence) has yet
been reported. It was further suggested that the combination of the
DM measurements of FRBs and the redshift measurements of GRBs could
open a new window to study cosmology (i.e., to probe the history of
the free electron column density and thus the cosmic reionization;
Deng \& Zhang 2014; Gao et al. 2014; Zhou et al. 2014). This attempt
could be feasible and effective if the DM of FRBs is indeed
overwhelmingly determined by the combination of the IGM and the
Galaxy. However, conservatively speaking, some uncertainties could
still arise because some substantial DM contributions could be
provided from somewhere else.

The most probable intrinsic DM contributor for GRB-associated FRBs
could be the GRB ejecta, whose contribution was estimated (but
somewhat underestimated) by Deng \& Zhang (2014) by using the usual
DM definition for stationary medium. In fact, for the
relativisticaly moving ejecta, relativistic transformation should be
taken into account in the DM calculations. More importantly, the DM
of an FRB could also be contributed by the wind blown from the
precollapsing NS. The existence of the NS wind was evidenced by its
significant influence on the GRB afterglow emission either by
injecting energy into the GRB external shock (Dai \& Lu 1998a,
1998b; Zhang \& M{\'e}sz{\'a}ros 2001) or by producing internal
emission through energy dissipation of the wind (Troja et al. 2007;
Mao et al. 2010; Yu et al. 2010). The efficiency of the wind
emission depends on the specific dissipation mechanisms and, more
directly, the amount of electrons carried by the wind. As an
intuitive consideration, more electrons are probably required by
brighter wind emission, which then could determine a higher DM for
the corresponding FRB.

Therefore, the primary purpose of this paper is (1) to clarify the
possible DM contributions to FRBs from GRB ejecta and NS winds and
(2) to reveal the consequent implications for the cosmological
application of the expected FRB-GRB association. In the next
section, I give a relativistic definition of the DM. The DM
contribution from the GRB ejecta is estimated in Section 3, where
the dynamical influence on the GRB external shock by the energy
injection effect is taken into account. In Section 4, I pay
attention to the DM contribution from the NS wind by considering two
different lepton-loading cases. The conclusion and discussion are
given in Section 5.

\section{Dispersion in relativistic moving material}
It is widely accepted that both GRB ejecta and NS winds move at
ultra-relativistic speeds (e.g., Lithwick \& Sari 2001; Zhang et al.
2003; Zou \& Piran 2010; Coroniti 1990; Lyubarsky \& Kirk 2001;
Drenkhahn 2002; Metzger et al. 2008). Thus, the usual definition of
DM for stationary medium as the column density of electrons could
become inappropriate for GRB ejecta and NS winds. Alternatively, the
Lorentz transformation between the observer's frame and the comoving
frame needs to be taken into account. Throughout this paper, the
comoving quantities are labeled by a superscripted prime.

When an electromagnetic wave propagates through an ionized medium of
electron number density $n'$, the group velocity of the wave would
become frequency-dependent, which reads $v'_{\rm emw}(\nu')={c
[1+{\nu'_{\rm p}}^2/({\nu'}^2-{\nu'_{\rm p}}^2)]^{-1/2}}$, where
$\nu'_{\rm p} = (n'e^2/\pi m_{\rm e})^{1/2} = 9 \times 10^3
{n'}^{1/2}$ Hz is the plasma frequency (Rybicki \& Lightman 1979).
Therefore, after propagating through the medium, the arrival of the
electromagnetic wave of frequency $\nu'\gg \nu'_{\rm p}$ should be
delayed by a time $\Delta t' \approx (1/c)\int ({\nu'}_{\rm
p}^2/2{\nu'}^2)dl'$ with respect to the arrival time in vacuum.
Then, in the observer's frame, the arrival time delay of a radio
pulse between two frequencies $\nu_1$ and $\nu_2$ can be calculated
using
\begin{eqnarray}
\Delta t &=& {(1+z)\Delta t'\over \mathcal D}\nonumber \\
&\approx& \frac{e^2}{2\pi m_{\rm e} c}
\left(\frac{1}{\nu_{1}^2}-\frac{1}{\nu_{2}^2}\right)\int \frac
{\mathcal D}{1+z}n'dl'\nonumber\\
&\equiv& \frac{e^2}{2\pi m_{\rm e} c}
\left(\frac{1}{\nu_{1}^2}-\frac{1}{\nu_{2}^2}\right){\rm DM}
\label{DM-def0},
\end{eqnarray}
where $z$ is the cosmological redshift of the medium and $\mathcal
D$ is the Doppler factor due to the relativistic speed $v$ of the
medium. As usual I denote $\beta=v/c$, the Lorentz factor
$\Gamma=(1-\beta^2)^{-1/2}$, and $\mathcal D=[\Gamma(1-\beta
\cos\theta)]^{-1}$ with $\theta$ being the angle between the line of
sight and the direction of medium motion. In Equation
(\ref{DM-def0}), the DM of the radio pulse measured by the observer
is defined as
\begin{equation}
{\rm DM}\equiv\int \frac {\mathcal D}{1+z}ndl, \label{DM-def}
\end{equation}
where the Lorentz invariance of the column density (i.e.,
$n'dl'=ndl$) is considered. In comparison with the usual DM
definition, two extra quantities appear in Equation (\ref{DM-def}),
i.e., the Doppler factor $\mathcal D$ and redshift $z$.

\section{DM contributed by GRB ejecta}
Following Huang et al. (1999, 2000), the dynamical evolution of a
GRB ejecta, which propagates into the surrounding medium and shocks
it, can be determined from the energy conservation law as follows:
\begin{eqnarray}
E(t)=\left(\Gamma_{\rm ej}-1\right)M_{\rm ej}c^2+\left(\Gamma_{\rm
sm}^2-1\right)M_{\rm sm}c^2,\label{dyn1}
\end{eqnarray}
where $\Gamma_{\rm ej}$ and $M_{\rm ej}$ ($\Gamma_{\rm sm}$ and
$M_{\rm sm}$) are the Lorentz factor and the mass of the GRB ejecta
(the shocked medium), respectively. Moreover, we have $\Gamma_{\rm
ej}=\Gamma_{\rm sm}$ and $M_{\rm sm}=4\pi r^{3}n m_{\rm p}/(3-k)$,
where $r$ is the radius of the GRB external shock from the central
engine. As usual, the density of the circumburst medium can be
written as $n=Ar^{-k}\rm cm^{-3}$ with the index $k=0$ and 2
corresponding to the interstellar medium and main-sequence stellar
wind environments, respectively (Dai \& Lu 1998c; Chevalier \& Li
2000).

By considering of the possible energy injection from an NS wind to
the GRB ejecta and shocked medium, the temporal dependence of the
isotropically equivalent total energy can be written as
\begin{eqnarray}
E(t)=E_{0}+\dot{E}_{\rm w}t/(1+z),\label{dyn2}
\end{eqnarray}
where $E_{0}$ is the initial energy of the ejecta impulsively
released during the GRB, $\dot{E}_{\rm w}$ is the energy flux of the
NS wind, and the time $t$ is measured in the observer's frame. GRB
afterglow observations implied that GRB NSs are millisecond
magnetars, i.e., with a polar magnetic field of $B_{\rm
p}\sim10^{14}$ G and a spin period of $P\sim1$ ms (Zhang et al.
2006; Fan \& Xu 2006; Yu \& Dai 2007; Yu et al. 2010; Rowlinson et
al. 2013; Gompertz et al. 2013). Then the wind energy flux
determined by the spin-down luminosity of the NS can be estimated
as\footnote{By considering different structures of the NS
magnetosphere, this estimation can be corrected by a factor of the
order of unity (e.g., Spitkovsky 2006; Li et al. 2012).}
$\dot{E}_{\rm w}=L_{\rm sd}/f_{\rm b}\approx{B_{\rm p}^2R^6\Omega^4/
(6c^3f_{\rm b})}=10^{48}B_{\rm p,14}^{2}P_{-3}^{-4}R_{6}^6f_{\rm
b,-1}^{-1}~\rm erg~s^{-1}$, where $R$ and $\Omega$ are the radius
and angular frequency of the NS, respectively, and the factor
$f_{\rm b}$ is introduced due to the possible beaming of the NS
wind. Hereafter the conventional notation $Q_{x}=Q/10^{x}$ is
adopted in cgs units.

Combining Equations (\ref{dyn1}) and (\ref{dyn2}), the Lorentz
factor of the ejecta can be approximately expressed as
\begin{eqnarray}
\Gamma_{\rm ej}(t) &\approx&\left\{
\begin{array}{ll}
\eta,&{\rm for}~t<t_{\rm dec},\\
\left({E_0\over M_{\rm sm}c^2}\right)^{1/2},&{\rm for}~t>t_{\rm dec},\\
\left[{\dot{E}_{\rm w}t\over (1+z)M_{\rm sm}c^2}\right]^{1/2},&{\rm
for}~t>t_{\rm ei},
\end{array}\right.
\end{eqnarray}
where $\eta=E_0/ M_{\rm ej}c^2$ represents the initial value of
$\Gamma_{\rm ej}$, $t_{\rm dec}$ is the deceleration timescale
determined by the condition $M_{\rm sm}=M_{\rm ej}/\eta$, and
$t_{\rm ei}\equiv{(1+z)E_0/\dot{E}_{\rm w}} $ is the time at which
the injected energy starts to influence the dynamics by exceeding
the initial energy of the ejecta. Obviously, the relative importance
of the energy injection effect depends on the competition between
the initial energy $E_{0}$ and the total injected energy
$\dot{E}_{\rm w}t_{\rm col}/(1+z)$, where $t_{\rm col}$ is the NS
collapsing time. More strictly, it is probable that, actually, only
a fraction (sometimes a small fraction) of the wind energy can be
injected into the external shock, because the other fraction is
emitted directly by the wind itself to produce the internal
afterglow emission (see Section 4.2). In any case, for a typical
value of $t_{\rm col}\sim10^3$ s indicated by the observed internal
plateaus, the total injected energy can be estimated to be at most
$\sim10^{51}$ erg.

If an FRB is produced by an NS collapse at the observational time
$t$, the radius of the GRB ejecta, where the FRB crosses it, can be
approximately calculated by
\begin{eqnarray}
r_{\rm ej,c}={cv_{\rm ej}t/(1+z)\over c-v_{\rm ej}}\approx{
2\Gamma_{\rm ej}^2ct\over(1+z)}.
\end{eqnarray}
Therefore, the DM of the FRB contributed by the GRB ejecta can be
calculated as
\begin{eqnarray}
{\rm DM}_{\rm ej}={ 2\Gamma_{\rm ej} \over (1+z)}{M_{\rm
ej}\over4\pi r_{\rm ej,c}^2 m_{\rm p}},
\end{eqnarray}
where $\mathcal D_{\rm ej}\approx 2\Gamma_{\rm ej}$ is taken for
$\theta\approx0^\circ$. On one hand, for a mild GRB explosion with
$E_0\ll10^{51}$ erg, the dynamical evolution of the ejecta should be
taken as $\Gamma_{\rm ej}=[{\dot{E}_{\rm w}t/ (1+z)M_{\rm
sm}c^2}]^{1/2}$ and $\Gamma_{\rm ej}\propto t^{-(2-k)/(8-2k)}$. This
yields
\begin{eqnarray}
{\rm DM}_{\rm
ej}^{k=0}&=&0.1(1+z)^{1/4}\dot{E}_{\rm w,48}^{-3/8}E_{0,50}^{}\eta_{2.5}^{-1}n_{0}^{3/8}t_{3}^{-5/4}~{\rm cm^{-3}pc},\\
{\rm DM}_{\rm ej}^{k=2}&=&8(1+z)^{}\dot{E}_{\rm
w,48}^{-3/4}E_{50}^{}\eta_{2.5}^{-1}A_{35.5}^{3/4}t_{3}^{-2}~{\rm
cm^{-3}pc}.
\end{eqnarray}
As shown, the DM contribution from the GRB ejecta here is
insignificant in contrast to the IGM's contribution, although the
energy of the ejecta is finally increased. This is because the
low-energy ejecta has too few electrons. On the other hand, if the
GRB explosion is very powerful with $E_0\gg10^{51}$ erg, which
meanwhile indicates a high-mass ejecta, the ``standard" dynamics
$\Gamma_{\rm ej}=({E_0/ M_{\rm sm}c^2})^{1/2}$ and $\Gamma_{\rm
ej}\propto t^{-(3-k)/(8-2k)}$ should be taken for the ejecta. Then
we can get
\begin{eqnarray}
{\rm DM}_{\rm
ej}^{k=0}&=&5(1+z)^{-1/8}E_{0,52}^{5/8}\eta_{2.5}^{-1}n_{0}^{3/8}t_{3}^{-7/8}~{\rm cm^{-3}pc},\\
{\rm DM}_{\rm
ej}^{k=2}&=&150(1+z)^{1/4}E_{52}^{1/4}\eta_{2.5}^{-1}A_{35.5}^{3/4}t_{3}^{-5/4}~{\rm
cm^{-3}pc}.
\end{eqnarray}
In comparison with Deng \& Zhang (2014), the above values have
obviously increased due to the relativistic correction. In
particular, in the case of the main-sequence stellar wind
environment, the GRB ejecta could provide a substantial contribution
to the DM of the corresponding FRB. Finally, in all cases, the
plasma frequency of the GRB ejecta can be found to be safely lower
than the radio frequency ($\sim 10^9$ Hz).

In addition, the DM contribution from a shocked medium with much
fewer electrons than the ejecta can definitely be neglected.

\section{DM contributed by neutron star wind}
The corotating magnetosphere of an NS is filled with electron and
positron pairs (Goldreich \& Julian 1969). Beyond the light
cylindrical radius $r_{\rm L}=c/\Omega$, the corotation can no
longer hold and the magnetocentrifugal force exerted on the pairs
would throw them with relativistic speed. Therefore, it is widely
considered that an NS wind probably carries a certain amount of
leptons, while the energy of the wind is initially dominated by
Poynting flux. In this section the DM contribution of such a
lepton-loaded NS wind is assessed, including cases (1) the leptons
are only provided by the NS magnetosphere and (2) a great amount of
leptons are provided from somewhere else other than the
magnetosphere. In the sight of afterglow emission, these two types
of NS winds correspond to a shallow-decay (or a normal) afterglow
and an internal plateau afterglow, respectively. The former one is
emitted from the GRB external shock that is energized by the NS
wind, while the latter one is produced by the NS wind self.

\subsection{Goldreich-Julian Wind}
The leptons carried by an NS wind can at least be provided by the NS
magnetosphere, where the particle density can be expressed as
$n_{\rm GJ}(r)\approx{(\Omega B_{\rm p}/ 2\pi c
e)}\left({r/R}\right)^{-3}$ (Goldreich \& Julian 1969; Shapiro \&
Teukolsky 1983). Here the angle-dependence of the density is ignored
for simplicity. Moreover, the NS wind is considered to be
approximately isotropic initially and gradually becomes collimated
far away from the star (denoted by a beaming factor $f_{\rm b}$).
Then the particle number flux of the NS wind can be calculated by
\begin{eqnarray}
\dot{N}_{\rm GJ}&\approx&4\pi r_{\rm L}^2 n_{\rm GJ}(r_{\rm
L})c/f_{\rm b}\nonumber\\
&=&5.5\times10^{39}B_{\rm p,14}^{}P_{-3}^{-2}R_{6}^3f_{\rm
b,-1}^{-1}~\rm s^{-1}.\label{number}
\end{eqnarray}
After the collapse of the NS, the energy supply to the NS wind is
turned off and the remnant wind material expands outside quickly.
Therefore, the FRB can cross the Goldreich-Julian (GJ) particles
only when it catches up with the wind at the radius $r_{\rm
w,c}\approx 2\Gamma_{\rm w}^2 r_{\rm L}$, where $\Gamma_{\rm w}$ is
the Lorentz factor corresponding to the bulk motion of the wind. The
wind velocity is considered to have a radial direction due to
large-scale acceleration and collimation, although it is initially
dominated by the tangential component at $r_{\rm L}$.

The process of wind acceleration at large radii is uncertain, but it
is widely considered that magnetic reconnections could play an
important role in it. Following Lyubarsky \& Kirk (2001) and
Drenkhahn (2002), magnetic reconnection acceleration could determine
the dynamical evolution to be $\Gamma_{\rm w}(r)\sim\Gamma_{\rm
L}\left(r/ r_{\rm L}\right)^{\alpha}$, where the index $\alpha$
could be within the range of $1/3-1/2$. The initial speed of the
wind at the light cylinder is set to the Alfv{\'e}n speed, and the
corresponding Lorentz factor reads $\Gamma_{\rm
L}\sim\sqrt{\sigma_{\rm L}}$ with $\sigma_{\rm L}$ representing the
initial ratio between the Poynting flux and the matter energy flux
(Drenkhahn 2002). Therefore, from the expression $\dot{E}_{\rm
w}=(\sigma_{\rm L}+1)\Gamma_{\rm L}\dot{N}_{\rm GJ}m_{\rm e}c^2$, we
can derive $\Gamma_{\rm L}\sim({\dot{E}_{\rm w}/ \dot{N}_{\rm
GJ}m_{\rm e}c^2})^{1/3}=6\times10^4B_{\rm
p,14}^{1/3}P_{-3}^{-2/3}R_{6}$. Due to such a high initial Lorentz
factor and subsequent acceleration, the crossing radius $r_{\rm
w,c}$ can easily be larger than the radius of the GRB ejecta. This
means that the wind material would merge into the GRB ejecta before
it is caught up by the FRB. After the merger of the wind and ejecta,
their DM contributions can simply be compared through their lepton
numbers as $M_{\rm ej}/m_{\rm p}=E_{0}/\eta m_{\rm
p}c^2\sim2.2\times10^{51}E_{0,51}\eta_{2.5}^{-1}$ and $\dot{N}_{\rm
GJ}t/(1+z)\sim5.5\times10^{42} (1+z)^{-1}B_{\rm
p,14}^{}P_{-3}^{-2}R_{6}^3f_{\rm b,-1}^{-1}t_{3}$. Obviously, the DM
contribution of the GJ particles is negligible.

\subsection{Internally-Emitting Wind}
The most substantial evidence for a remnant GRB NS is the observed
internal X-ray afterglows, which exhibit a plateau followed by an
extremely steep decay (Troja et al. 2007; Liang et al. 2007;
Rowlinson et al. 2010, 2013), typically with a luminosity of $L_{\rm
X}\sim 10^{46-47}\rm erg~s^{-1}$ in the X-ray band. As far as I can
see, no similar temporal behavior has been found in the optical and
high-energy afterglows. The extremely steep decay probably indicates
the collapse of the NS. Therefore, Zhang (2014) suggested that the
steep decay following an internal plateau could be associated with
an FRB signal that is produced by the NS collapse.

Although our knowledge of the internal dissipation mechanism of NS
winds is very limited, an intuitive idea could arise that strongly
emitting NS winds could carry much more leptons than the GJ flux.
Thus, here I take the particle number flux $\dot{N}_{\rm IE}$ as a
free parameter. Then the emission luminosity of a wind, most of
which is assumed to enter into the X-ray band according to the
present observations, can be written as
\begin{eqnarray}
L_{\rm X}&\sim&\dot{N}_{\rm IE}\Gamma_{\rm w}\gamma' m_{\rm e}
c^2,\label{LX}
\end{eqnarray}
where $\Gamma_{\rm w}$ and $\gamma'$ are the bulk Lorentz factor of
the wind and the comoving random Lorentz factor of electrons,
respectively. Furthermore, the wind emission is assumed to be
dominated by synchrotron radiation, i.e., (Sari et al. 1998)
\begin{eqnarray}
(1+z)\nu_{\rm X}&\sim&\Gamma_{\rm w}{e B'{\gamma'}^2\over 2\pi
m_{\rm e}c},\label{nuX}
\end{eqnarray}
where $\nu_{\rm X}\sim3\times10^{17}$ Hz is taken as a reference
frequency and $B'$ is the comoving strength of the magnetic field in
the NS wind. By introducing an equipartition factor $\epsilon$, we
can write
\begin{eqnarray}
{{B'}^2\over 8\pi}&\sim&\epsilon{\dot{N}_{\rm IE}\gamma' m_e
c^2\over 4\pi r_{\rm e}^2 \Gamma_{\rm w} c},\label{Bs}
\end{eqnarray}
where the emission is considered to mainly happen at the radius
$r_{\rm e}$.

Solving Equations (\ref{LX}$-$\ref{Bs}), we can obtain the emission
radius as follows:
\begin{eqnarray}
r_{\rm e}&\sim&{e{\gamma'}^2\over \pi m_{\rm
e}c^{3/2}(1+z)}\left({\epsilon L_{\rm X}\over 2\nu_{\rm
X}^2}\right)^{1/2}\nonumber\\&=&2.3\times
10^7(1+z)^{-1}{\gamma'}^2\epsilon^{1/2}L_{\rm X,47}^{1/2}\nu_{\rm
X,17.5}^{-1}~\rm cm.\label{re}
\end{eqnarray}
Obviously, the emission radius of the wind should not be larger than
the simultaneous radius of the GRB external shock ($\sim10^{16-17}$
cm). Such a requirement gives
${\gamma'}\lesssim3\times10^4(1+z)^{1/2}\epsilon^{-1/4}L_{\rm
X,47}^{-1/4}\nu_{\rm X,17.5}^{1/2}$. On the other hand, the optical
depth of the wind at the emission radius reads
\begin{eqnarray}
\tau&=&\sigma_{\rm T}{\dot{N}_{\rm IE}t/(1+z)\over 4\pi r_{\rm
e}^2}\nonumber\\
&\sim&1.2\times10^{16}(1+z)\Gamma_{\rm
w}^{-1}{\gamma'}^{-5}\epsilon^{-1}\nu_{\rm X,17.5}^2t_{3},
\end{eqnarray}
where $\sigma_{\rm T}$ is the Thomson cross section. In order to be
consistent with the non-thermal assumption, the optical depth is
required to be much smaller than unity, which yields $ \gamma'>
480(1+z)^{1/6}a^{-1/6}\epsilon^{-1/6}\nu_{\rm
X,17.5}^{1/3}t_3^{1/6}$ with $a\equiv\Gamma_{\rm w}/ \gamma'$. The
value of the parameter $a$ depends on the specific dissipation
mechanisms of the wind.

With the derived range of $\gamma'$, the internal-emission-required
electron flux
\begin{eqnarray}
\dot{N}_{\rm IE}\sim{{L_{\rm X}\over \Gamma_{\rm w}{\gamma'} m_{\rm
e} c^2}}=1.2\times{10^{53}\Gamma_{\rm w}^{-1}{\gamma'}^{-1}}L_{\rm
X,47}~\rm s^{-1}\label{Ne}
\end{eqnarray}
can be roughly constrained to be within the range of
\begin{eqnarray}
1.4\times10^{44}a^{-1}{\rm s^{-1}}\lesssim \dot{N}_{\rm
IE}<5.3\times 10^{47}a^{-2/3}{\rm s^{-1}},
\end{eqnarray}
where the relatively certain parameters are omitted for clarity. The
above result is drastically larger than the GJ flux presented in
Equation (\ref{number}), although an uncertainty still exists due to
the uncertain parameter $a$. It is at least demonstrated that, in
order to produce the bright internal plateau afterglows, a great
number of electron-positron pairs must be generated and accelerated
somewhere from the light cylinder to the emission radius (i.e.,
$r_{\rm L}<r_{\rm e^{\pm}}<r_{\rm e}$).

The DM contributed by the spontaneously generated electron-positron
pairs can be calculated from
\begin{eqnarray}
{\rm DM_{w}}&\approx&{2\Gamma_{\rm w}\over (1+z)}{\dot{N}_{\rm IE}
t/(1+z)\over 4\pi r_{\rm
w,c}^2}\nonumber\\&\sim&{1.6\times10^{36}L_{\rm X,47}t_{3}\over
(1+z)^2\Gamma_{\rm w}^{4}{\gamma'}r_{\rm e^{\pm}}^{2}}~\rm
cm^{-3}pc,\label{DM2}
\end{eqnarray}
where $r_{\rm w,c}=2\Gamma_{\rm w}^2r_{\rm e^{\pm}}$. If the DM of
FRBs is mainly attributed to the IGM, then we should require ${\rm
DM_{w}}\ll 10^3~\rm cm^{-3}pc$, which constrains the pair-generation
radius to be $r_{\rm e^{\pm}}\gg
1.3\times10^{9}(1+z)^{-1}a_0^{-2}{\gamma'}_{3}^{-5/2}L_{\rm
X,47}^{1/2}t_{3}^{1/2}\rm ~cm$. Such a condition does not seem
unreachable, since after all the radius $r_{\rm e^{\pm}}$ is at
least larger than the radius of the NS. In contrast, if the
pair-generation radius is indeed small, it could become possible
that the DMs of some FRBs are actually dominated by NS winds.
Nevertheless, a high DM usually corresponds to a high plasma
frequency, which should of course be lower than the radio frequency,
i.e.,
\begin{eqnarray}
\nu_{\rm p}&=&{\Gamma_{\rm w}\over 1+z}\left({e^2\over \pi m_{\rm
e}}{\dot{N}_{\rm
IE}\over 4\pi r_{\rm w,c}^2\Gamma_{\rm w} c}\right)^{1/2}\nonumber\\
&=&{2.6\times10^{24}L_{\rm X,47}^{1/2}\over (1+z)\Gamma_{\rm
w}^{2}{\gamma'}^{1/2}r_{\rm e^{\pm}}}{\rm Hz}\label{nup2}\ll10^9{\rm
Hz}.
\end{eqnarray}
This requires $r_{\rm e^{\pm}}\gg
8.1\times10^{7}(1+z)^{-1}a_0^{-2}{\gamma'}_{3}^{-5/2}L_{\rm
X,47}^{1/2}\rm ~cm$. Otherwise, the FRB signal would be absorbed by
the wind plasma. Therefore, for
$10^{8}a_0^{-2}{\gamma'}_{3}^{-5/2}{\rm ~cm}\lesssim r_{\rm
e^{\pm}}\lesssim10^{9}a_0^{-2}{\gamma'}_{3}^{-5/2}\rm ~cm$, the wind
material could contribute an extremely high DM to the corresponding
FRB, which exceeds the IGM's contribution.

\section{Conclusion and discussion}
Since a supermassive NS could form during some GRBs and live for
hundreds and thousands of seconds after the birth, the GRBs are
expected to be associated with an FRB which is produced by the
collapse of the NS. (1) GRB observations showed that the wind of
some GRB NSs could produce bright internal afterglow emission, which
indicates that a great number of electron-positron pairs are
generated and accelerated beyond the light cylinder of the NS. If
the pair-generation radius is small enough, these leptons could
contribute a high DM to the FRB signal, which can be comparable to
and even exceed the DM contribution from the IGM. (2) If the leptons
carried by the NS wind are only supplied by the NS magnetosphere
(i.e., for a GJ wind), the wind emission is probably very weak and
most wind energy will be injected into the GRB external shock. In
such a case, the FRB could be associated with a shallow-decay
afterglow rather than an internal plateau. Meanwhile, the DM of the
FRB could be overwhelmingly contributed by a combination of the IGM
and the Galaxy. (3) Finally, in the GJ-wind case, the FRB could also
be associated with a normal single-power-law afterglow, if the
impulsive energy released during the GRB is larger than the total
injected energy. The huge prompt energy indicates a high mass of the
ejecta, so a DM of the order of $\sim100\rm cm^{-3}pc$ could be
predicted for a circumburst environment of main-sequency stellar
wind. Such a situation could appear in some long-duration GRBs.

In summary, on one hand, we must be very careful to assess the
various possible contributions to the FRB's DM when we use the
FRB-GRB association as a cosmological probe. On the other hand, the
DM measurements of GRB-associated FRBs could provide a constraint on
the physics of NS winds.

In principle, by considering of various possible DM contributors,
the Galactic origin of FRBs cannot be ruled out before their
cosmological redshifts are measured. For example, Loeb et al. (2014)
recently proposed that FRBs could be rare eruptions of flaring
main-sequence stars within $\sim$1 kpc, where the high DMs of the
FRBs arise from a blanket of coronal plasma around the host stars.
Following a similar consideration, in the NS scenarios, a high DM
could also be contributed by the relativistic NS wind with some
peculiar properties (e.g. with a millisecond period and a normal
magnetic field of $10^{11-12}$ G). Of course, some difficulties
could be argued against such intrinsic DM origin models, e.g., the
density of the plasma could be too high to enable the penetration of
the radio emission (Luan \& Goldreich 2014; Tuntsov 2014; Dennison
2014). In any case, it could be valuable to deeply investigate the
dynamical evolution of a steady NS wind (rather than the remnant
wind discussed in this paper) to judge whether the Galactic NSs
could produce FRBs or not.

\acknowledgements The author thanks the Hong Kong University of
Science and Technology for hospitality while this work was being
completed and appreciate the helpful discussions with B. Zhang and
K. S. Cheng. This work is supported by the 973 program (grant No.
2014CB845800), the National Natural Science Foundation of China
(grant Nos. 11103004 and 11473008), and the Program for New Century
Excellent Talents in University (grant No. NCET-13-0822).

\end{document}